\documentclass[fleqn,usenatbib]{mnras}

\usepackage[T1]{fontenc}
\usepackage{ae,aecompl}

\usepackage{graphicx}	
\usepackage{amsmath}	
\usepackage{amssymb}	

\usepackage{color}

\newcommand{\Mpc}            {\,{\rm Mpc}}
\newcommand{\Msun}           {\,{\rm M}_\odot}

\newcommand{\rh}            {r_{\rm h}}
\newcommand{\Mstr}          {M_{\rm star}}

\newcommand{\tinf}          {t_{\rm infall}}

\newcommand{\subfind}       {{\small SUBFIND}}
\newcommand{\fig}           {Fig.~\ref}

\definecolor{purple}{rgb}{0.5, 0., 0.5}

\newcommand{\red}[1]{\textcolor{black}{#1}}

\title[A tale of two populations]{A tale of two populations: surviving and destroyed dwarf galaxies and the build-up of the Milky~Way's stellar halo}

\author[A. Fattahi et al.]{Azadeh Fattahi$^{1}$\thanks{E-mail: azadeh.fattahi-savadjani@durham.ac.uk}, Alis J. Deason$^{1}$, Carlos S. Frenk$^{1}$, Christine M. Simpson$^{2,3}$, \newauthor Facundo A. G\'{o}mez$^{4,5}$, Robert J. J. Grand$^{6}$, Antonela Monachesi$^{4,5}$, Federico \newauthor Marinacci$^7$, R\"udiger Pakmor$^6$\\\\
$^{1}$Institute for Computational Cosmology, Department of Physics, University of Durham, South Road, Durham DH1 3LE, UK\\
$^{2}$Enrico Fermi Institute, The University of Chicago, Chicago, IL 60637, USA\\
$^{3}$Department of Astronomy \& Astrophysics, The University of Chicago, Chicago, IL 60637, USA\\
$^{4}$Instituto de Investigaci\'on Multidisciplinar en Ciencia y Tecnolog\'ia, Universidad de La Serena, Ra\'ul Bitr\'an 1305, La Serena, Chile\\
$^{5}$Departamento de Astronom\'ia, Universidad de La Serena, Av. Juan Cisternas 1200 Norte, La Serena, Chile\\
$^{6}$Max-Planck-Institut f\"{u}r Astrophysik, Karl-Schwarzschild-Str. 1, 85748 Garching, Germany\\
$^{7}$Department of Physics \& Astronomy, University of Bologna, via Gobetti 93/2, 40129 Bologna, Italy
}

\date{Accepted XXX. Received YYY; in original form ZZZ}

\pubyear{2020}

\begin{document}
\label{firstpage}
\pagerange{\pageref{firstpage}--\pageref{lastpage}}
\maketitle

\begin{abstract}
  We use magneto-hydrodynamical simulations of Milky Way-mass haloes
  from the Auriga project to investigate the properties of surviving
  and destroyed dwarf galaxies that are accreted by these haloes over
  cosmic time. We show that the {\it combined} luminosity function of
  surviving and destroyed dwarfs at infall is similar in the various
  Auriga haloes, and is dominated by the destroyed dwarfs. There is,
  however, a strong dependence on infall time: destroyed dwarfs
  typically have early infall times of less than $6$~Gyr \red{(since the
  Big Bang)}, whereas the majority of dwarfs accreted after $10$~Gyr
  have survived to the present day. Because of their late infall, the
  surviving satellites have higher metallicities at infall than their
  destroyed counterparts of similar mass at infall; the difference is
  even more pronounced for the present-day metallicities of
  satellites, many of which continue to form stars after infall, \red{in
  particular for $\Mstr>10^7 \Msun$}. In agreement with previous work,
  we find that a small number of relatively massive destroyed dwarf
  galaxies dominate the mass of stellar haloes. However, there is a
  significant radial dependence: while 90 percent of the mass in the
  inner regions ($<\,20\,$kpc) is contributed, on average, by only 3
  massive progenitors, the outer regions ($>\,100\,$kpc) typically
  have $\sim8$ main progenitors of relatively lower mass.  Finally, we
  show that a few massive progenitors dominate the metallicity
  distribution of accreted stars, even at the metal-poor end. Contrary
  to common assumptions in the literature, stars from dwarf galaxies
  of mass $\Mstr<10^7 \Msun$ make up less than 10 percent of the
  accreted, metal poor stars ([Fe/H] $<\,-3$) in the inner $50\,$kpc.

\end{abstract}




\section{Introduction}

The hierarchical nature of galaxy formation in the $\Lambda$CDM cosmological model implies that galaxies are surrounded by a diffuse stellar halo formed by the accretion and disruption
of lower mass galaxies \citep[e.g.][]{Bullock2005,Cooper2010}. This formation mechanism indicates that the halo is composed of relatively old, metal-poor stars, many still part of substructures associated with accretion events. The early discovery of the Sagittarius stream \citep{Newberg2002, Majewski2003} and \red{Helmi stream \citep{Helmi1999b}} around the Milky Way, and later of the field of streams \citep{Belokurov2006}, as well as observations of nearby galaxies such as the Andromeda galaxy \citep{McConnachie2009}, strongly favour this overall picture for the formation of stellar haloes.

The Galactic halo, which, because of its proximity, can be resolved into individual stars and substructures, provides a unique window into the assembly history of the Milky Way (MW), and allows important tests of the
galaxy formation framework. Recent surveys, in particular the \textit{Gaia} mission, are revolutionising our understanding of the formation and evolution of our Galaxy, including its stellar halo, by providing 6D phase-space information and chemical data for a large number of stars \citep{Gaia_DR1,Gaia_DR2}. For example, the discovery of the Gaia-Sausage-Enceladus population of stars in highly eccentric orbits 
uncovered a significant accretion event in the history of the MW \citep{Belokurov2018, Haywood2018, Helmi2018, Myeong2018}\footnote{Note, however, that there is some debate as to whether or not Gaia-sausage and Gaia-Enceladus refer to the same structure and event; see, e.g. \citet{Evans2020,Elias2020}.}. 
This population is thought to have been brought in by a relatively massive 
dwarf galaxy whose destruction generated a significant fraction of the inner stellar halo 
\citep{Fattahi2019,Mackereth2019,Bignone2019}; this is consistent with predictions from 
cosmological simulations \citep{Bullock2005,Cooper2010}. 

Before evidence of the hierarchical formation of the stellar halo
  became available, a commonly held view was that the halo had formed
  by `monolithic collapse' \citep{Eggen1962}. Early arguments against
  formation through accretion included the apparent lack of extremely
  metal-poor stars in dwarf galaxies and the different [$\alpha$/Fe]
  abundance patterns of the stellar halo and dwarf galaxies
  \citep[see, e.g.,][]{Gilmore1998,Helmi2006,Tolstoy2009}. However,
  improving observational techniques in deriving metallicities and the
  discovery of very low metallicity stars in Sculptor and other dwarfs, addressed the first objection
  \citep{Kirby2008,Frebel2010,Starkenburg2013b}. Moreover, it was recognized
  that the observed difference in the [$\alpha$/Fe] abundance patterns
  could be explained if the stellar halo formed from the early
  accretion of relatively massive dwarf galaxies
  \citep{Robertson2005,Font2006}. A second debate ensued as to
  whether the building blocks of the halo are analogueous to the dwarfs
  that survive at the present or whether they are fundamentally
  different kinds of galaxies
  \citep[e.g][]{Tolstoy2003,Venn2004}. This is one of the main topics
  that we address in this work.  Previous discussions of this topic
  may be found in
  \citet{Read2006a,Purcell2007,Sales2007a,Tissera2012,Chua2017,Fiorentino2017}.

The basic features of galactic halo formation were established in
N-body cosmological simulations of MW-mass haloes combined with simple
models for the stellar component \citep[such as particle tagging
methods;][]{Bullock2005,Cooper2010}.  In particular, these models
showed that the mass of galactic stellar haloes is dominated by a few
massive destroyed dwarf galaxies and that the contribution from
low-mass and ultra-faint dwarfs is negligible
\red{\citep{Deason2015,Deason2016,Fiorentino2017,D'souza2018,Amorisco2017,Monachesi2019}}. High-resolution
hydrodynamical simulations of MW-mass haloes in $\Lambda$CDM, where
the dwarf galaxies responsible for the stellar halo are resolved,
became possible in the past few years \citep[e.g., APOSTLE, Auriga,
Latte, \red{`ELVIS on
  Fire'};][]{Sawala2016b,Fattahi2016,Grand2017,Wetzel2016,Garrison-Kimmel2018},
leading to more detailed predictions and interpretation of the
observational data. These simulations offer the opportunity to follow
the formation of stellar haloes in a realistic and self-consistent way
by including processes not modelled in N-body simulations, such as gas
physics, star formation after infall, self-consistent metallicities,
and the formation of an {\it in-situ} component of the stellar halo.

The in-situ halo component seen in hydrodynamical simulations consists
of kinematically hot stars born in the main progenitor of the galaxy.
The three main mechanisms for the formation of this kinematically hot
in-situ component are: (i) heated disc stars, (ii) stars formed in the
halo from cooling gas, (iii) stars formed from stripped gas from
accreted dwarfs \citep{Cooper2015}.  The fraction of these stars and
the contribution from the various formation channels are simulation
dependent and therefore highly \red{debated \citep[see
  also,][]{Font2011b,Zolotov2009,Pillepich2015,Pillepich2018,Monachesi2016}. }
Indeed, \citet{Monachesi2019} show that the mass of stellar haloes in
the Auriga simulations are in better agreement with observations when
only the accreted component is considered.

\red{In this work, we analyze the Auriga suite of cosmological
  hydrodynamical simulations of MW-mass haloes. Our main goal is to
  compare the overall properties of dwarf galaxies that have been
  tidally disrupted and whose stars make up the accreted galactic
  stellar halo with the properties of the dwarf galaxies that survive
  as satellites. We then focus on the destroyed population and examine
  the radial assembly and metallicity build-up of stellar haloes. This
  latter part of the paper is complementary to the work of
  \citet{Monachesi2019} who studied the assembly history and general
  properties of galactic stellar haloes in the Auriga simulations. We
  extend that work to include an analysis of the radial dependence of
  the assembly history of the stellar halo, as well as the
  implications for the build-up of the dark matter halo and for the
  most metal-poor component of the stellar halo.}

This paper is organized as the following; Sec.~\ref{sec:sim} describes the simulations; Sec.~\ref{sec:two_populations} compares the luminosity function, infall time, metallicity and gas content of destroyed and surviving satellites; 
Sec.~\ref{sec:buildup} describes the assembly of Auriga stellar haloes, and  Sec.~\ref{sec:halometals} the metallicity build-up of the halo as  a function of radius. We conclude with a summary in Sec.~\ref{sec:summary}.


\begin{figure*}
	\includegraphics[width=17.cm]{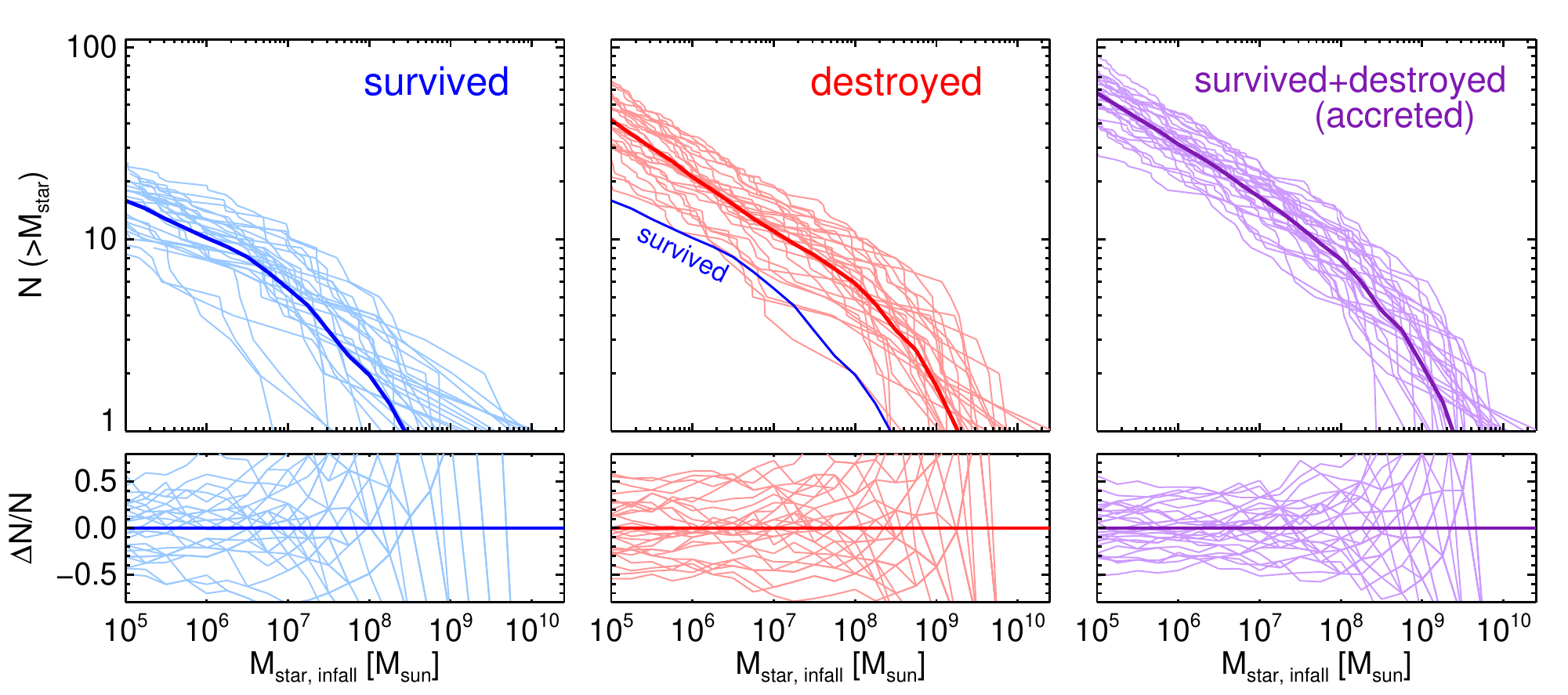}
	\caption{\red{{\it Top row:} stellar mass function (SMF) at
            infall of dwarf satellites that survive to $z=0$ (left),
            destroyed dwarfs (middle) and the combination of both
            (right). Thin curves show individual Auriga haloes, and
            thick darker curves the average (average number at fixed
            stellar mass) amongst the 28 haloes. The labelled blue
            line in the middle panel shows the average SMF of
            surviving dwarfs, repeated from the left panel. {\it
              Bottom row}: scatter around the average of the stellar
            mass functions in the top row. The rms around the mean SMF
            at the low mass end ($<10^7 \Msun$) is $\Delta N/N=0.32$,
            $0.31$, and $0.24$ for satellites, destroyed dwarfs, and
            combined populations, respectively. The SMF of the
            combined surviving and destroyed dwarf populations has
            less scatter than the individual SMFs.}}
    \label{fig:SMF}
\end{figure*}

\begin{figure*}
	\includegraphics[width=17.cm]{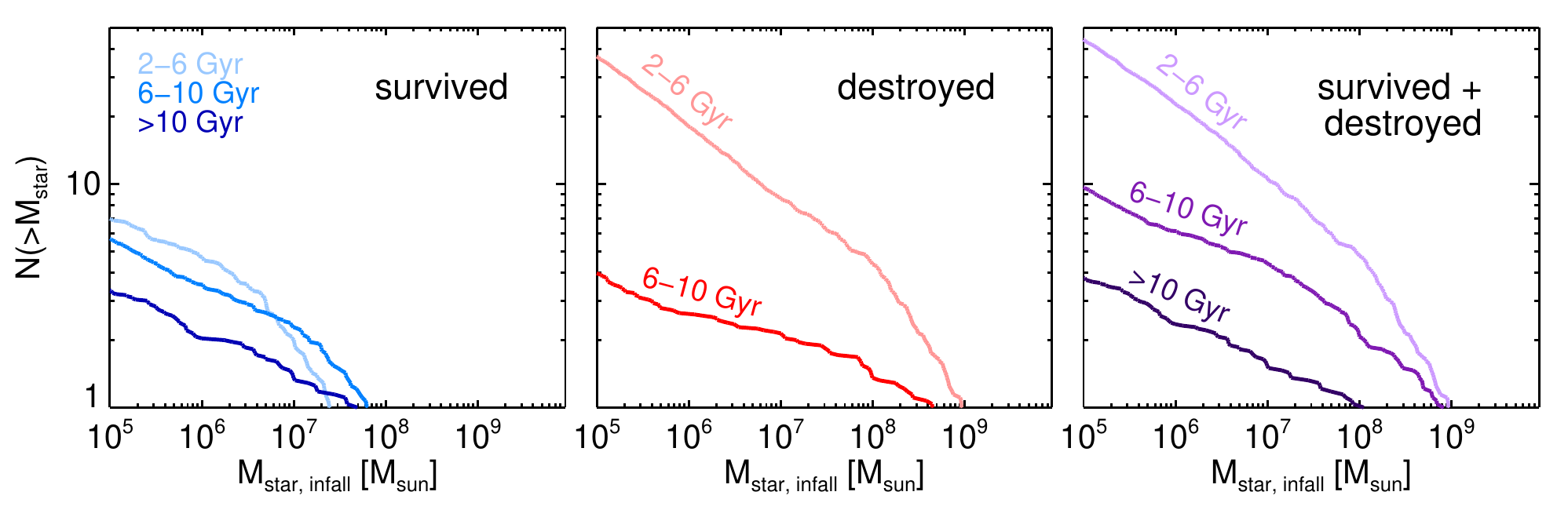}
	\caption{Similar to \fig{fig:SMF} but now the stellar mass
          functions (SMFs) are divided into three infall time bins, 
          $\tinf=2-6\,$Gyr, $\tinf=6-10\,$Gyr, and $\tinf>10\,$Gyr
          \red{(present day corresponds to 13.8 Gyr)}, as indicated in the legend. The
          curves represent the average SMF over 28 Auriga haloes. The
          SMFs strongly depend on infall time, and the typical infall
          times of the surviving and destroyed populations are very
          different.}
    \label{fig:SMF_t}
\end{figure*}

\section{Auriga Simulations}
\label{sec:sim}

In this study we use cosmological, magneto-hydrodynamical (MHD)
simulations of MW-mass haloes from the Auriga project
\citep{Grand2017}. The Auriga suite consists of `zoom-in' simulations
\citep{Frenk1996} of relatively isolated haloes with virial masses of
$M_{200} \sim 10^{12}\Msun$, which were chosen from the $100^3 \Mpc^3$
periodic cube of the EAGLE project \citep{Schaye2015,Crain2015}. The
simulations start at $z=127$ from initial conditions made by
Panphasia \citep{Jenkins2013}, and are developed to $z=0$ by the
Tree-PM, moving-mesh code, {\sc arepo}
\citep{Springel2011,Weinberger2019}. The subgrid galaxy formation
model is described in detail in \citet{Grand2017} and
\citet{Marinacci2014}. In summary, it includes metal line cooling,
star formation, stellar evolution feedback, supermassive black hole
formation and feedback, homogeneous UV photoionizing radiation with
reionization redshift $z_{\rm re}=6.5$. The simulations assume 
cosmological parameters in accordance to \citet{Planck2015}:
$\Omega_{\rm m}=0.307$, $\Omega_{\Lambda}=0.693$,
$\Omega_{\rm bar}=0.048$, and a Hubble parameter of $h=0.6777$.

We analyze the original 30 Auriga haloes of mass,
$M_{200}= (1-2) \times 10^{12}\Msun$, at the fiducial level (L4)
resolution, $m_{\rm DM}\sim3\times 10^5 \Msun$, with gas particle mass
resolution, $m_{\rm bar}\sim5\times10^4\Msun$, and a maximum Plummer
equivalent gravitational softening, $\epsilon_{\rm max}=369$~pc. Six
of the Auriga haloes have been simulated at the higher resolution
level (L3), at which, $m_{\rm DM}\sim4\times 10^4 \Msun$,
$m_{\rm bar}\sim6\times10^3\Msun$ and $\epsilon_{\rm max}=184$ pc.

Dark matter haloes in the simulations were identified using a
Friends-of-Friends (FoF) algorithm \citep{Davis1985} and bound
structures and substructures within FoF groups were found iteratively
using \subfind\,\citep{Springel2005b}. MW analogues refer to the
central subhalo (subhalo-0) of the main FoF groups.  \citet{Grand2017}
present an analysis of the galactic discs in the simulations and show
that they reproduce the general properties of disc-dominated
galaxies. \citet{Simpson2018} shows that the luminosity function of
the dwarf satellites in Auriga matches that of the Milky Way
satellites. Furthermore, the sizes and star formation histories of the
dwarf satellites have also been shown to agree with observations
\citep{Bose2019,Digby2019}.

In this work, we use merger trees to track galaxies in time only after
$z=3.1$ ($t=2.1$ Gyr \red{since the Big Bang}), due to the uncertainties in the
identification of the main progenitors at earlier times. We refer to
destroyed dwarfs as those that have fallen into the main
host\footnote{Infall is defined as crossing the $r_{200}$ radius of the main
  halo for the first time. Infall parameters are based on the snapshot
  immediately before crossing $r_{200}$.} more recently than $z=3.1$ and
retain no bound remnant (according to \subfind) at the present time. For
the fiducial resolution (L4), being destroyed is equivalent to galaxy
masses falling below $M_{\rm DM}\sim 10^{7} \Msun$ and
$M_{\rm star}\sim 10^5
\Msun$. 
Satellites are identified as surviving bound substructures within
$r_{200}$ of the main haloes at $z=0$. Stellar mass and average
metallicities of dwarfs at any given time are defined based on the bound
star particles inside twice their 3D stellar half mass radius
($\rh$). The reference frames of the MW analogues are based on \subfind,
i.e. the position of the particle with the minimum gravitational
potential. The orientations of galactic discs are defined according to
the angular momentum of stars within 10~kpc. 

\begin{figure}
    \hspace{-0.3cm}
	\includegraphics[width=8.7cm]{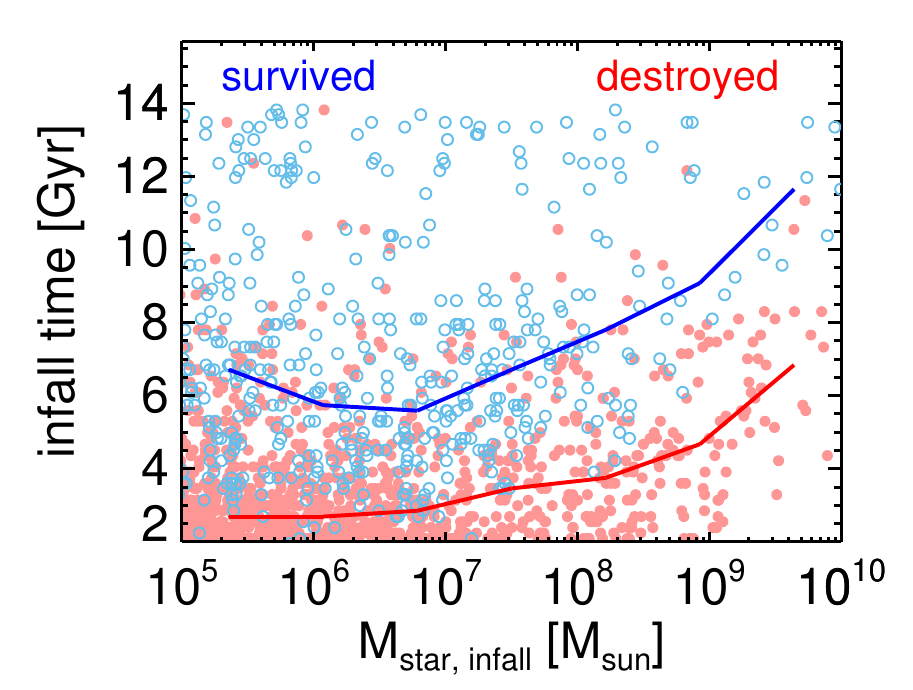}\\	 
	\caption{Infall time \red{($t=0$ corresponds to the Big Bang)}
          vs. stellar mass at infall for destroyed (red) and surviving
          (blue) dwarfs accreted onto the 28 Auriga haloes. The
          curves of the corresponding colour show the average infall time at a
          fixed stellar mass. There is a clear distinction between the
          infall times of surviving and destroyed dwarfs. }
    \label{fig:tinfall}
\end{figure}

\begin{figure}
    \hspace{-0.2cm}
	\includegraphics[width=8.2cm]{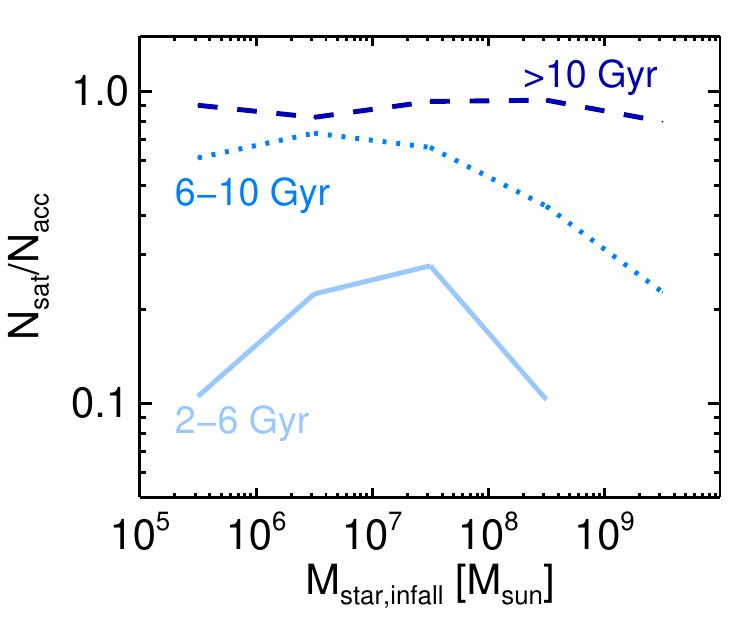}
	\caption{Fraction of surviving dwarfs (satellites) relative to
          all dwarfs that are accreted onto the Auriga haloes, as a
          function of stellar mass at infall. Different curves
          correspond to the infall time bins of 
          Fig.~\ref{fig:SMF_t}. Almost all dwarfs accreted at late
          times survive to the present day. At earlier times, the
          fraction of dwarfs that survives depends on the mass of the
          dwarf: more massive dwarfs are subject to dynamical
          friction, and a more rapid destruction, but lower mass
          dwarfs are more susceptible to tidal effects.}
    \label{fig:fraction}
\end{figure}

The results presented in this work include 28 Auriga haloes, rather
than the full suite of 30, since we discard two of them (Au-11 and
Au-20) as they are undergoing a merger at $z=0$.

\subsection{Accreted and in-situ forming stars}

In this study, accreted or ex-situ\footnote{We use the terms ex-situ
  and accreted interchangeably throughout this paper.} stars are
defined as those that are bound to the MW analogues (not to 
satellites) at $z=0$, \red{but were born in a halo or subhalo other
  than the main progenitor of the MW. In practice, the birth location
  is established at the snapshot immediately following the formation
  of the star.} Since we do not track galaxies before $z=3.1$,
stars that are formed before that redshift are flagged as accreted or
in-situ according to their membership at $z=3.1$. Such old stars make
up $\sim 4$ percent of the final stellar mass in Auriga galaxies and
therefore do not affect our results significantly.

According to our definition of accreted and in-situ stars, stars that
are formed in satellite galaxies before and \textit{after} infall are
flagged as accreted, while stars that are formed out of stripped gas
from satellites (i.e. gas that is no longer bound to the satellites) are
flagged as in-situ. Indeed, \citet{Cooper2015}, using a suite of three hydrodynamical simulations of MW-mass haloes, found that most of the gas stripped from satellites
is accreted onto the disc of the galaxy and forms stars there.

\section{Destroyed vs. surviving dwarf galaxies}
\label{sec:two_populations}

We present the cumulative stellar mass function (SMF) at infall of 
surviving satellites\footnote{We use the terms satellites and
  surviving satellites interchangeably in this work.}, destroyed
dwarfs, and the combination of the two populations (i.e. all accreted)
in
the top row of \fig{fig:SMF}. Lighter colour curves show 
individual haloes, while the thick solid lines show the average of all
curves in each panel.
This figure includes dwarf galaxies with only a few star particles. We
discuss convergence using the L3 simulations in
Appendix~\ref{sec:convergence}, and show that the results are very
well converged. We emphasize that the total accreted SMF (rightmost
panel) does not suffer from potential numerical artifacts related to
tidal stripping and disruption of subhaloes
\citep{vdB2018,Errani2020}, since those effects might, if anything,
change the relative number of destroyed and survived dwarfs, but not
the total number of accreted ones.

This total accreted (i.e. destroyed+survived) SMF is the outcome of
the dark matter subhalo accretion history combined with the stellar
mass-halo mass relation at different redshifts. The former is a
prediction of the $\Lambda$CDM structure formation model and has been
shown to be relatively similar amongst haloes of similar mass
\citep{GuoWhite2008,Fakhouri2010,Ludlow2013,JiangCole2015}. The
stellar mass-halo mass relation depends on the galaxy formation model
of the simulations, but has little scatter at fixed stellar mass, at
the low mass end \citep{Simpson2018}. Therefore, it is not surprising
that the \textit{overall} accreted SMF also has relatively small scatter
amongst various haloes.

Fig.~\ref{fig:SMF} shows that there are fewer satellites, on average,
relative to destroyed dwarfs; \red{28 percent of all accreted
  dwarfs with infall mass $\Mstr>10^5 \Msun$, and 33 percent of those with $\Mstr > 10^7 \Msun$
  survive to $z=0$.} However, this does not necessarily imply that the
mass of the accreted stellar halo in MW-mass galaxies is larger than
the combined stellar mass of the surviving satellites. For example,
not all of the accreted mass will end up in the stellar halo; a
fraction will end up in the disc and a small fraction in the bulge
\citep{Gomez2017,Gargiulo2019}. We checked that in these Auriga
haloes the median ratio between the accreted mass inside $r_{200}$ and
the combined stellar mass of the surviving satellites at $z=0$ is
$M_{\rm acc}(<r_{200})/\Sigma M_{\rm sat} = 3.4$, with a large scatter
of 1.0 dex. This ratio changes to
$M_{\rm acc}(<r_{200})/\Sigma M_{\rm sat} = 1.2$ when
considering the accreted component \textit{outside} the disc region
($|z|>5$ kpc and $R>20$ kpc). Thus, the definition of ``stellar halo''
is an important consideration when comparing the total mass of
accreted components with observations.


\red{It is interesting to note that the halo-to-halo variation in the stellar mass of the very brightest destroyed dwarf galaxies in Fig.~\ref{fig:SMF} is nearly a factor of 100, 
  even though the masses of MW analogs are similar. Since it is the remnants of these bright dwarf galaxies that dominate the stellar haloes, this variation explains some of the observed diversity in the stellar haloes of  MW-mass galaxies \citep[see, also][]{Deason2016,D'souza2018,Monachesi2019}. We discuss this further in Sec.~\ref{sec:buildup}.}

\begin{figure*}
	\includegraphics[width=17.cm]{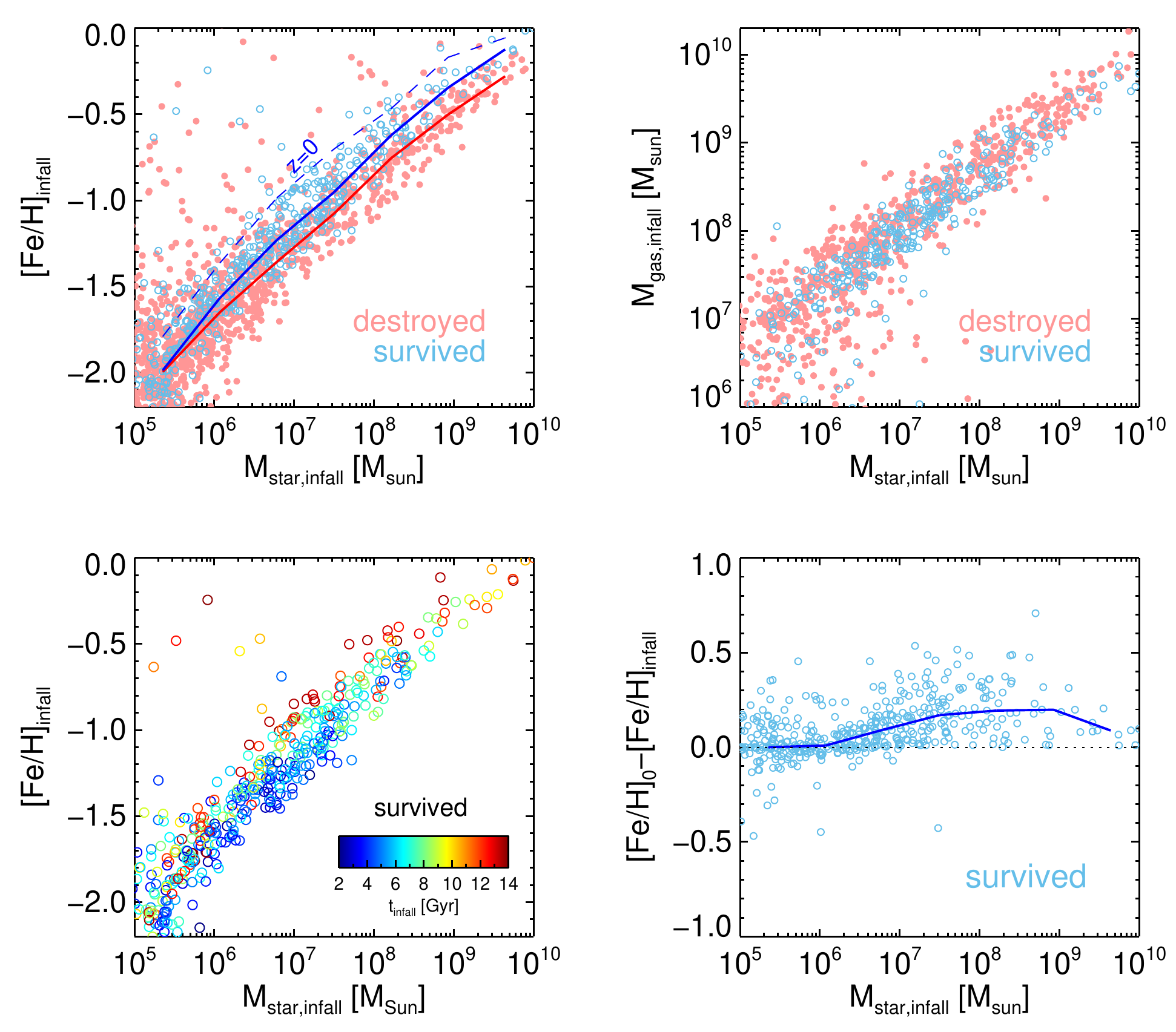}
	\caption{{\it Top-left}: the average stellar [Fe/H]
          vs. stellar mass, at infall for destroyed (red symbols) and
          surviving (blue symbols) dwarfs. The solid curves show the
          average [Fe/H] at fixed stellar mass. The dashed line shows
          the average $M_{\rm star}-$[Fe/H] relation of satellites at
          $z=0$. {\it Top-right}: total gaseous mass (inside
          $2\times\rh$) vs. stellar mass at infall for destroyed and
          surviving dwarfs; colours are similar to the top-left panel.
          {\it Bottom-left}: similar to top-left, but only for
          surviving satellites, colour coded according to 
          infall time, indicated by the colour bar, \red{as time after the Big Bang}. {\it
            Bottom-right}: the change in average [Fe/H] between infall
          and $z=0$ for surviving satellites.}
    \label{fig:z_mstar}
\end{figure*}

\subsection{Dependence on Infall time}

The SMF of satellites and destroyed dwarfs accreted over cosmic time
have a similar shape (but different normalisation). How does this
change as a function of infall time? \fig{fig:SMF_t} presents the SMFs
for the same three populations as in the previous figure, but divided
into three bins of infall time\footnote{Infall times throughout this
  work refer to the age of the Universe when the infall happens,
  not to lookback time; i.e. $\tinf=0$ is the Big Bang and 13.8 Gyr corresponds to present day.} (and
averaged over the 28 Auriga haloes). Firstly, it is worth noting the
relatively small variation in the SMF of surviving satellites to z=0 for
different infall times, in contrast to the destroyed population SMF
which shows a strong dependence on infall time\footnote{Note that only
  a few destroyed dwarfs over all the haloes, fell in after
  $\tinf=10\,$Gyr, hence there is no corresponding line in the middle
  panel.}; in particular, the population is dominated by early infall
dwarfs. Additionally, we note that the total accreted dwarfs are
dominated, in terms of number and mass, by these early infall dwarfs
($\tinf=2-6$ Gyr) which are mostly destroyed by the present time.

The bias in the infall time of the destroyed dwarfs compared to
survived satellites, as a function of stellar mass, is shown more
clearly in \fig{fig:tinfall}, where circles correspond to individual
dwarf galaxies in the 28 Auriga haloes, and the lines represent the
average as a function of infall stellar mass. Destroyed dwarfs fell
in, on average, $\sim5\,$Gyr earlier than satellites, with very few
($\sim$1 percent) falling after $t=10\,$Gyr; by contrast a
significant fraction ($\sim20$ percent) of surviving satellites have crossed the
virial radius of the host for the first time after
$t=10\,$Gyr. Additionally, we note that the infall times of accreted
dwarfs (either destroyed or surviving) increase with stellar
mass. \red{At a stellar mass of $M_{\rm star}\sim10^6 \Msun$,
  the typical infall times for destroyed and surviving dwarfs are
  $\tinf \sim 2.5$ Gyr and $\sim 6$ Gyr, respectively; at stellar mass of
  $M_{\rm star}\sim10^9 \Msun$ these numbers
  change to $\tinf \sim 5$ Gyr and $\sim 10$ Gyr for destroyed and
  surviving dwarfs, respectively.} This trend with mass reflects 
the hierarchical nature of galaxy formation in $\Lambda$CDM , where more massive
galaxies form later from the accretion and mergers of smaller
constituents. \red{We checked that these results are robust to changes
  in numerical resolution (see, Fig.~\ref{fig:tinfall_conv}).}

The survivability of accreted dwarfs depends both on their infall time
and their infall mass. We examine this in Fig.~\ref{fig:fraction}
where we illustrate the fraction of surviving satellites at $z=0$ relative
to the total number of accreted dwarfs,
$N_{\rm sat}/(N_{\rm sat}+N_{\rm destroyed})$, as a function of
infall stellar mass, and divided into different infall time bins. The
curves show averages over all the 28 haloes. As expected, a large
fraction of accreted dwarfs ($\sim 90$ percent) which fell in after
$\tinf>10$ Gyr survive to $z=0$, whereas only $10-20$ percent of the
early infall ($2 < \tinf/\mathrm{Gyr} < 6$) dwarfs survive as
classical satellites ($\Mstr>10^5\Msun$) today.

There are two interesting points about the trends with stellar mass in
Fig.~\ref{fig:fraction}. Firstly, more massive objects get destroyed
more efficiently at a given infall time. This is well understood to be the result of
dynamical friction, which strongly affects more massive objects and
drags their orbit towards the centre, hence leading to more
efficient tidal disruption. On the other hand, objects with lower stellar mass
appear to have a higher chance of getting disrupted. We have checked
these results in the L3 runs and confirm that this behaviour persists
at higher resolution (see Fig.~\ref{fig:fraction_conv}). Rather than
a resolution artifact, the behaviour at the low mass end can be
understood when considering the trend of infall time with mass (Fig.~\ref{fig:tinfall}), and the fact that lower mass objects have lower densities and are less resilient to tides. Interestingly, the combined effects of tides at the low mass end and dynamical friction at the high mass end, result in a mass range,
$M_{\rm star,infall}\sim10^7\Msun$, where accreted dwarf galaxies have the highest chance of survival (at a fixed infall time).

\subsection{Metallicity and gas content}

The different infall times of the two populations of destroyed and survived dwarfs results in
interesting differences in the properties of the destroyed dwarfs that
built up the stellar halo compared to the existing satellites. We
examine this in Fig.~\ref{fig:z_mstar}. The top-left panel, which
shows [Fe/H]\footnote{Defined as the median metallicity of star
  particles within $2 \times \rh$.} {\em vs} stellar mass at infall,
indicates that surviving satellites have higher metallicities at infall,
compared to their destroyed counterparts of similar
$M_{\rm star,infall}$. This is particularly evident at higher masses
\red{($\Delta$[Fe/H]$=0.2\,$dex at $M_{\rm
    star,infall}\sim10^9\Msun$)}. This trend can be understood by
considering the difference in the infall times of the two populations
and the evolution of the stellar mass-metallicity relation with
time. We see this in the bottom-left panel, where we show only
satellites and colour code them according to infall times. Higher mass
satellites fell in, on average, later and formed their stars over a longer a span of time
from pre-enriched gas, than lower mass satellites. Thus, the later
infall times of more massive satellite dwarfs leads them to have more
metal-rich stellar populations than their destroyed dwarf
counterparts, \red{which have typical infall times of $2-6$
  Gyr after the big~bang (Fig.~\ref{fig:tinfall})}. 

The dashed line in the top left panel shows the \textit{z=0}
  average stellar mass-[Fe/H] relation for surviving satellites. The
  difference compared to the relation prior to infall is twofold. The
  combination of stellar mass loss (moving left in this panel) and the
  increase of [Fe/H] after infall (see below) enhances the offset
  between the pre-infall stellar mass-metallicity relation and that of
  present day satellites.

\begin{figure*}
    \hspace{-0.12cm}
	\includegraphics[width=16.3cm]{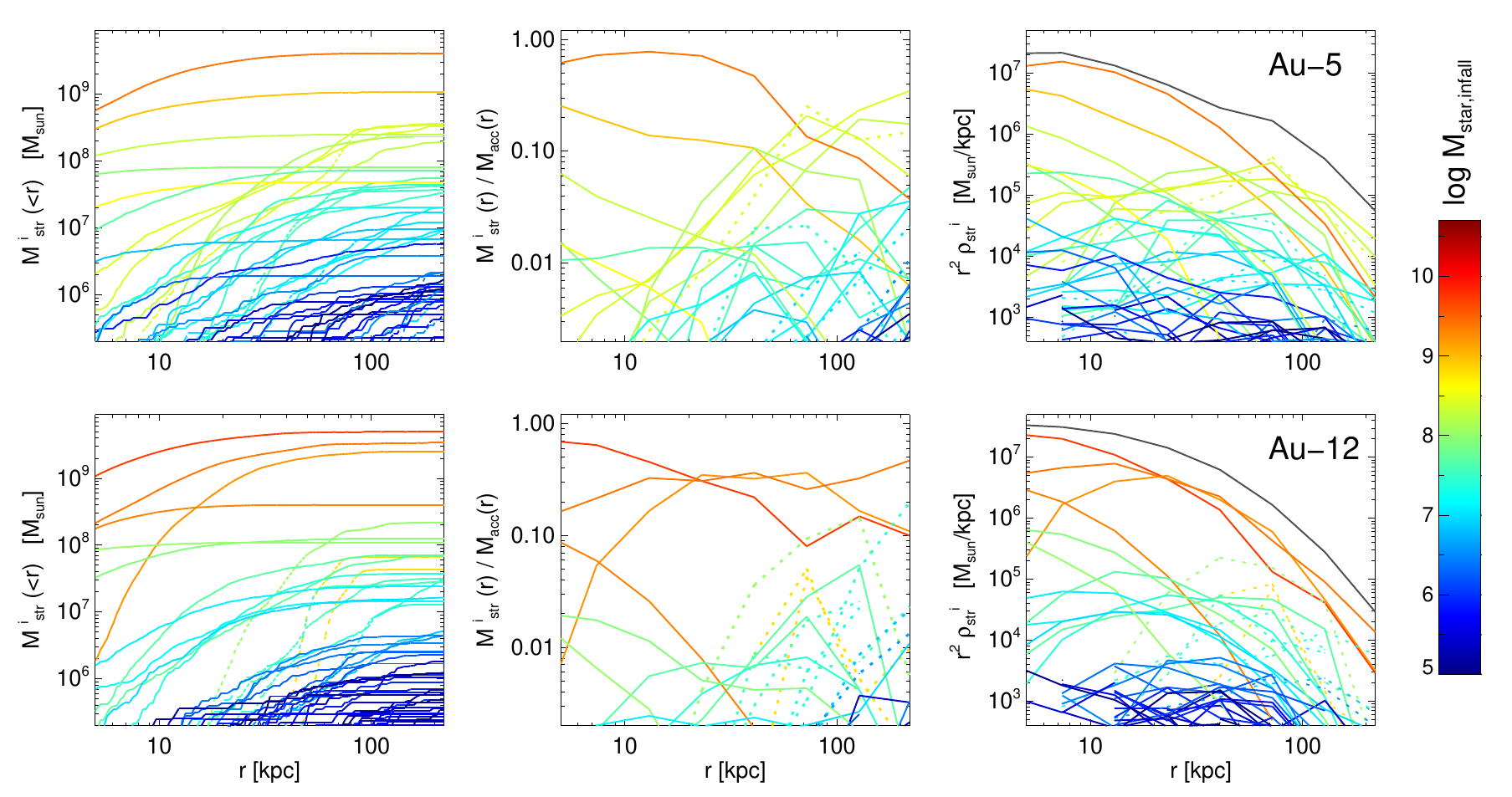}
	\caption{The radial distribution of all ex-situ stars originating 
          from different progenitor dwarf galaxies in Auriga halo 5
          (top row) and halo 12 (bottom row). Each curve corresponds
          to star particles associated to a single progenitor and is
          colour-coded according to the stellar mass at infall of the
          progenitor dwarf galaxy. Solid curves correspond to stars 
          originated from destroyed dwarfs, and dashed line to stars 
          stripped from surviving satellites. {\it Left}: enclosed mass profile
          of the stars from various progenitors. {\it Middle}:
          differential mass fraction (contribution) of stars from
          various progenitors, relative to the total ex-situ mass, in
          spherical shells. \red{{\it Right}:
            similar to the middle panel but showing the density profiles
            of various components ($\rho_{star}^{i} (r) \times r^2$),
            in spherical shells. The black curve shows the total
            profile of the accreted halo.} 
            The massive dwarf progenitors contribute
          the most stellar mass to the halo, particularly at low
          radii. However, the contribution from lower mass dwarfs
          becomes more significant beyond $\sim100\,$kpc} 
    \label{fig:radialdist}
\end{figure*}

All accreted dwarfs, but in particular the more massive ones, have a
significant amount of gas before infall, as shown in
Fig.~\ref{fig:z_mstar}; it is therefore not surprising that some
dwarfs keep forming stars after infall. The bottom-right panel of
Fig.~\ref{fig:z_mstar} shows that the average metallicity of high mass
satellites \red{($\Mstr>10^7\Msun$)} has increased after infall by
roughly 0.2~dex. This is due to their large gas (HI) reservoir,
combined with their ability to keep their cold gas for longer after
infall and form more stars. \citet{Simpson2018} presented a thorough
analysis of star formation after infall for Auriga satellites, and
show that higher mass dwarfs are more resilient to losing their gas
due to ram pressure stripping and keep forming stars after infall. At
the low mass end, the galaxies do not form many stars after infall,
and a larger fraction of them are gas poor even before
infall. \red{These results are consistent with the environmental effects on star formation associated with infall as observed in dwarf galaxies in the Local Group and nearby galaxies \citep{Geha2012,Slater2014,Fillingham2016}}.

\red{The decrease in the metallicities of a number of low mass dwarfs
  after infall, seen in the bottom right panel of
  Fig.~\ref{fig:z_mstar}, is likely due to insufficient numerical
  resolution combined with tidal stripping. These galaxies have a
  small number of star particles (less than $\sim 20$ at
  $\Mstr < 10^6 \Msun$) and the loss of just a few particles due to
  tidal stripping can change the metallicity noticeably. A comparison
  with higher resolution Auriga simulations is provided in the
  Appendix.}

Just as satellites, destroyed dwarfs would also have formed stars
after infall and before they were fully destroyed, and deposited stars
into the stellar halo. We find that these stars (i.e. formed after
infall of their progenitor) constitute only $10\pm5$ percent of the
mass of the accreted component, where the uncertainty is the {\em rms}
scatter amongst Auriga haloes.

The previous results, mainly the evolution of mass-metallicity relation with time, imply that galactic stellar haloes, which are
built up from destroyed dwarfs, are predicted to have a different
metallicity content than existing satellites.

\begin{figure*}
	\includegraphics[width=17.cm]{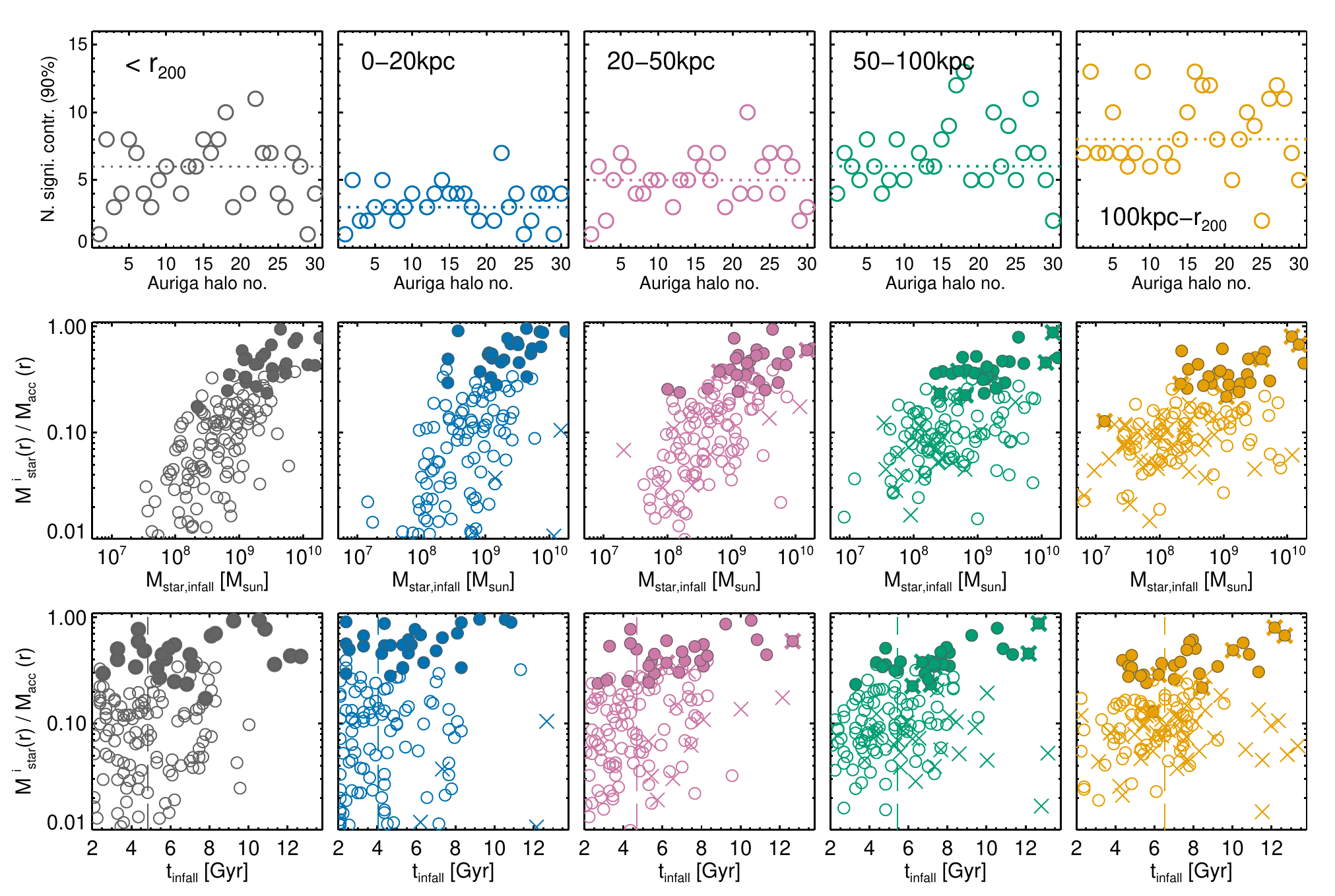}
	\caption{{\it Top row}: the number of significant progenitors
          at various radial (3D) bins, defined as the main
          contributors that make up 90 percent of the ex-situ mass in
          the given radial range. The leftmost panel includes all
          ex-situ stars within $r_{200}$ of the hosts, while the other
          panels correspond to various radial bins. The horizontal dotted line in each panel shows the median value over the 28 Auriga haloes. 
          {\it Middle row}: the mass contribution ($M_{i}(r)/M_{acc,tot}(r)$) of the top
          5 progenitors of the given radial range vs. the infall
          stellar mass of the progenitor dwarf galaxy. Circles and
          crosses indicates whether the progenitor is destroyed or has
          survived to $z=0$, respectively. Filled circles highlight
          the main contributor in each radial range,
          i.e. max($M_{i}(r)/M_{acc,tot}(r)$) for each Auriga
          halo. {\it Bottom row}: similar to the middle row but
          showing the infall times \red{($t=0$ corresponds to the Big Bang)} 
          of the top 5 progenitors. The vertical line in each panel shows the median infall time of these progenitors, over all haloes. The
          number of significant progenitors increases with radius, and
          the typical mass of these progenitors decreases with
          radius. Moreover, the inner regions of the halo are
          dominated by dwarfs accreted early, while the outer regions
          consists of material deposited at later times.}
    \label{fig:top5}
\end{figure*}

\begin{figure*}
    \includegraphics[width=17.cm]{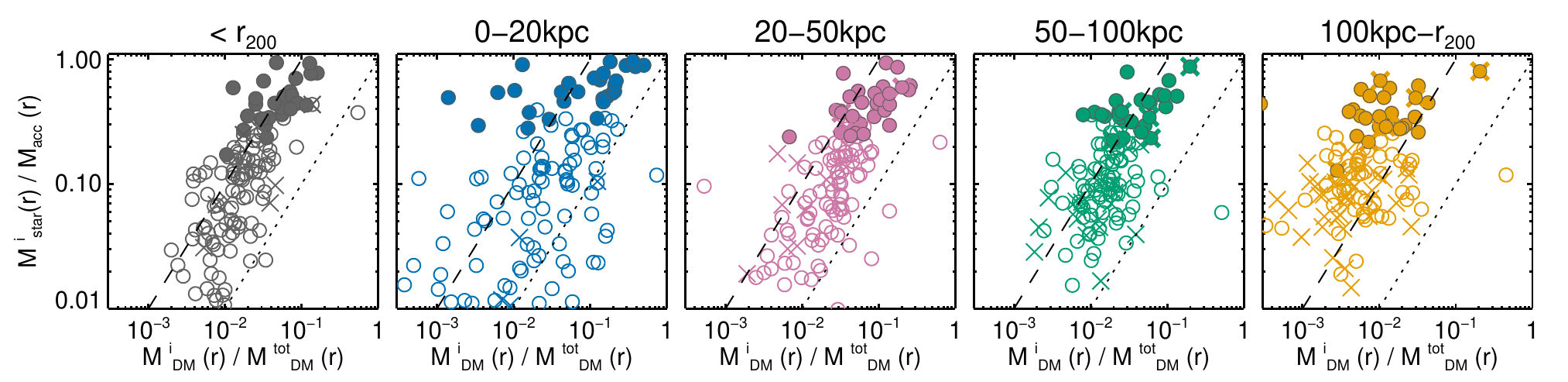}    
    \caption{The stellar mass contribution from the top 5 contributors
      to various radial ranges {\em vs} their dark matter contribution
      to the same region. The symbol types and panels are as in
      Fig.~\ref{fig:top5}. Dotted and dashed lines indicate constant
      1:1 and 1:10 ratios, respectively. The dark matter contribution
      from the main progenitors of the stellar halo are typically much
      smaller than their stellar contribution.}
    \label{fig:top5_dm}
\end{figure*}

\section{Build-up of the accreted stellar halo}
\label{sec:buildup}

In this section, we explore how the destroyed dwarf population builds
up the accreted stellar halo of MW-mass galaxies. \red{We extend the results of \citet{Monachesi2019} by focusing on the radial trends in the assembly of Auriga stellar haloes, as well as the metallicity and dark matter contribution of the destroyed dwarfs.} \red{We consider all accreted star particles inside $r_{200}$}. \red{We find that, on average, $\sim 15$ percent of these stars end up in the inner 5 kpc (roughly the bulge region), and $\sim 50$ percent in the disc region ($|z|<5$ kpc and $R<20$ kpc).}

Fig.~\ref{fig:radialdist} shows the stellar mass contribution, as a
function of radius, from all progenitors ($M_{\rm str}^i$) of the
accreted stellar mass for two example Auriga galaxies (top and bottom
rows). Each solid curve corresponds to an individual destroyed dwarf
galaxy; the dashed curves indicate debris from existing
satellites. The lines are all colour-coded by the progenitors'
stellar mass at infall. The enclosed mass profiles (left panels)
indicate that more massive dwarf galaxies contribute most to the total
mass of the stellar halo within $r_{200}$. Moreover, only a few
massive dwarf galaxies make up most of the accreted stars, and the
mass contribution from numerous low mass dwarfs is very small
\citep[see also,][]{Cooper2010,Monachesi2019}. However, \red{the middle} panels
of Fig.~\ref{fig:radialdist} show that the previous statement does {\it not}
hold at all radii, and that there is a notable radial dependence. These general trends are seen in all of the Auriga haloes (see
Fig.~\ref{fig:top5}). \red{The right panels of
  Fig.~\ref{fig:radialdist} show the density profile
  (i.e. $\rho_{star}^i (r) \times r^2$) of individual progenitors and
  confirm the conclusions of the previous panels.}  The two examples
in Fig.~\ref{fig:radialdist} are representative of cases where
a single dwarf galaxy dominates the build-up of the stellar halo (top row),
and where multiple massive dwarf galaxies form most of the halo mass
(bottom row).

Both examples in Fig.~\ref{fig:radialdist} show that the most massive
dwarfs typically deposit most of their stars in the inner regions
($<\,\sim50$\,kpc) and their contribution drops in the outer parts,
while lower mass dwarf galaxies deposit their stellar mass further out
and their contribution only becomes notable at $r>\,
\sim50\,$kpc. This behaviour is mainly due to dynamical friction,
which is stronger for more massive objects and causes their orbits to
sink to the centre of the haloes on a relatively short
timescale. Therefore, more massive objects are tidally stripped and
deposit their debris mainly in the inner regions. On the other hand,
the orbits of lower mass objects are less affected by dynamical
friction, but they are more susceptible to tidal disruption and their
debris is deposited along their orbits at relatively larger radii.


Fig.~\ref{fig:top5} summarises the previous results, but now extended
to the whole Auriga sample. The top row shows the number of
significant contributors to the stellar halo at various radii, defined
as the top-ranked contributors which formed 90 percent of the accreted
stellar mass in the given radial range. It is clear that the number of
significant progenitors increases with radius in all haloes; the inner
regions ($<20\,$kpc) are built, on average, from $3$ dwarf galaxies,
while the outer regions ($>100\,$kpc) have $\sim$8 significant progenitors. The
median number of significant contributors within $r_{200}$ (leftmost
panel) is $5$, \red{as also shown by \citet{Monachesi2019}}. 
The result within $r_{200}$ is mainly dominated by the
properties of the inner regions, as the stellar halo density decreases
with radius, so most of the mass is in the inner regions.

Do all significant progenitors contribute equally to the stellar halo,
and what are they properties? We address these questions in the middle
and bottom rows of Fig.~\ref{fig:top5}. Here, we show the top 5
progenitors of the given radial range for all halos, which are defined as those
that have contributed the most stellar mass to the given radial
range. As implied earlier, these typically build up 90 percent of the
inner stellar halo\footnote{At radii larger than 50 kpc, one could
  show $\sim10$ contributors which make up 90 percent of the halo;
  however, for clarity, we keep only the top 5 contributors.}. Circles
represent material originating in destroyed dwarf galaxies and the
crosses indicate debris from existing (surviving) dwarfs. The
contribution from this latter population is fairly minimal, but
becomes more important at larger radii (see below).

The middle row of Fig.~\ref{fig:top5} shows the stellar mass at infall of
the top 5 progenitors as a function of how much they contribute to the
accreted stellar mass in the given radial bin
($M_{\rm str}^{\rm i}(r)/M_{\rm str}^{\rm acc}(r)$). These panels
confirm our conclusions from the example haloes shown in
Fig.~\ref{fig:radialdist}: the inner regions ($<20\,$kpc) are strongly
dominated by very few relatively massive dwarf galaxies. Filled
circles highlight the top (main) contributor to each
radial bin for all Auriga galaxies, and they indicate that in most Auriga haloes \textit{only one} dwarf galaxy is enough to make up more than $\sim$50 percent of the accreted stellar mass in the inner regions. Moreover, there is a steep correlation
between the stellar mass of the progenitor dwarf and how much they 
contribute to the halo, such that the mass contribution from dwarf
galaxies less massive than $\Mstr<10^8 \Msun$ is negligible ($<1$
percent).

The outer parts of the halo ($>50\,$kpc) behave differently to the inner
regions. In most cases, there is no single progenitor that makes up
more than half of the halo and the contribution from various
progenitors becomes comparable. It is only in the outermost radial bin
($>100\,$kpc) that the contribution from $\Mstr<10^8 \Msun$ dwarfs is
non-negligible. Additionally, we can see that the debris from
surviving dwarfs (crosses) can significantly contribute to the outer
parts. On average $30 \pm 25$ percent of the mass in the
$100\,$kpc--$r_{200}$ radial range is made up of such debris. The
uncertainty range corresponds to the rms scatter. In comparison, the
debris fraction is negligible in the inner parts ($<1$ percent). It is
worth mentioning that only a small fraction of the total accreted mass
is in the outer regions, and the large \textit{fraction} of satellite
debris in the outskirts does not necessarily mean the majority of this
debris mass is in the outer parts.

The bottom row of Fig.~\ref{fig:top5} is similar to the middle row but shows the infall time of the top 5 progenitors in each radial range. We can clearly see the inside out formation of the stellar halo in these panels: the top 5 progenitors of the inner regions typically fell in before $\tinf=8$ Gyr, while infall times move towards the present as one considers larger radii. The median infall time of \textcolor{black}{these top 5 progenitors}, shown by the vertical lines, changes from $\tinf=4\,$Gyr ($z=1.6$) in the inner 20 kpc to $\tinf=6.5\,$Gyr ($z=0.8$) in the outer 100 kpc. \textcolor{black}{Considering only the main contributor (filled circles) the infall times move closer to the present time. Additionally, the radial} trend is less clear when one considers only the main contributor. This is because the main progenitors are more massive dwarfs (as can be seen from the middle row), and their mass is relatively dominant at all radii. Moreover, dynamical friction affects their orbits significantly and causes them to sink to the middle regardless of their infall time; hence the existence of late infall ($\tinf\sim10\,$Gyr) main progenitors in the innermost bin. 

\red{The diversity in the assembly of the stellar haloes is apparent
  in the left panels of Fig.~\ref{fig:top5}; while some haloes are
  built up primarily by one or two dwarf galaxies, some require 8-10
  dwarfs to account for 90 percent of their mass. Moreover, the
  contribution from the top progenitor of each halo (filled circles in
  the middle and bottom rows) can vary between 20 and almost 100
  percent. These top progenitors span almost two orders of magnitude
  in mass at infall, and have a large range of infall times. It is
  therefore not surprising that observations of nearby MW-mass
  galaxies, e.g. the GHOST survey, Dragonfly, and PAndAS, show a
  variety of properties in their stellar haloes
  \citep{Merritt2016,Monachesi2016a,Harmsen2017,McConnachie2009}. These results are
  consistent with studies using other simulations
  \citep{Cooper2010,Deason2016,D'souza2018}, and reconfirms the results from \citet{Monachesi2019} that used the Auriga simulations. The inside-out formation of the stellar halo and the radial trends shown in Fig.~\ref{fig:top5} have significant implications about the metallicity built-up of the stellar halo, which we discuss in Sec. \ref{sec:halometals}. }

\subsection{Implications for accreted dark matter}

The dwarf galaxies that build up the stellar halo, also contribute to the
dark matter halo of the galaxy. Due to the non-linear stellar
mass-halo mass relation and to differences in tidal stripping, the dark
matter dark matter contribution of these dwarf galaxies is expected
to be different from their stellar
contribution. 
For the top 5 contributors to the accreted stellar halo
(Fig.~\ref{fig:top5}), Fig.~\ref{fig:top5_dm} shows their dark matter mass
contribution at various radii
($M_{\rm dm}^{\rm i}(r)/M_{\rm dm}^{\rm tot}(r)$). This is calculated by
flagging dark matter particles that were bound to the dwarf galaxy progenitors
at infall, and, after disruption of the dwarf galaxy, are now (at
$z=0$) bound to the host.

Generally, the dark matter contribution of individual dwarfs is
significantly lower, by almost an order of magnitude, compared to
their stellar contribution. These results can be understood by
considering that, (i)~a considerable fraction of the dark matter mass
is built up by smooth accretion and the disruption of dark subhaloes
\citep{Fakhouri2010a,Genel2010}; and, (ii)~the steep stellar-halo mass
relation at low masses \citep{Simpson2018,Moster2013,Behroozi2013}
implies that dwarf galaxies over a large range of stellar mass
contribute similarly to the dark matter. Moreover, the results have a
larger scatter at each radial bin which is due to the difference in
the tidal stripping of dark matter and stars from an accreted dwarf;
stars are embedded deeply in the gravitational well of galaxies and
are resilient to tides, as opposed to extended dark matter haloes which
get stripped first in the outer parts of the halo. In other words,
most of the stars from massive dwarfs are deposited in the inner
regions while their dark matter is extended throughout the halo.
\red{The results from this section explain why the phase-space distribution of accreted stars is not directly linked to that of the total DM at a given Galactocentric distance, as pointed out by \citet{Bozorgnia2019}.}

\begin{figure}
    \hspace{-0.2cm}
	\includegraphics[width=8.4cm]{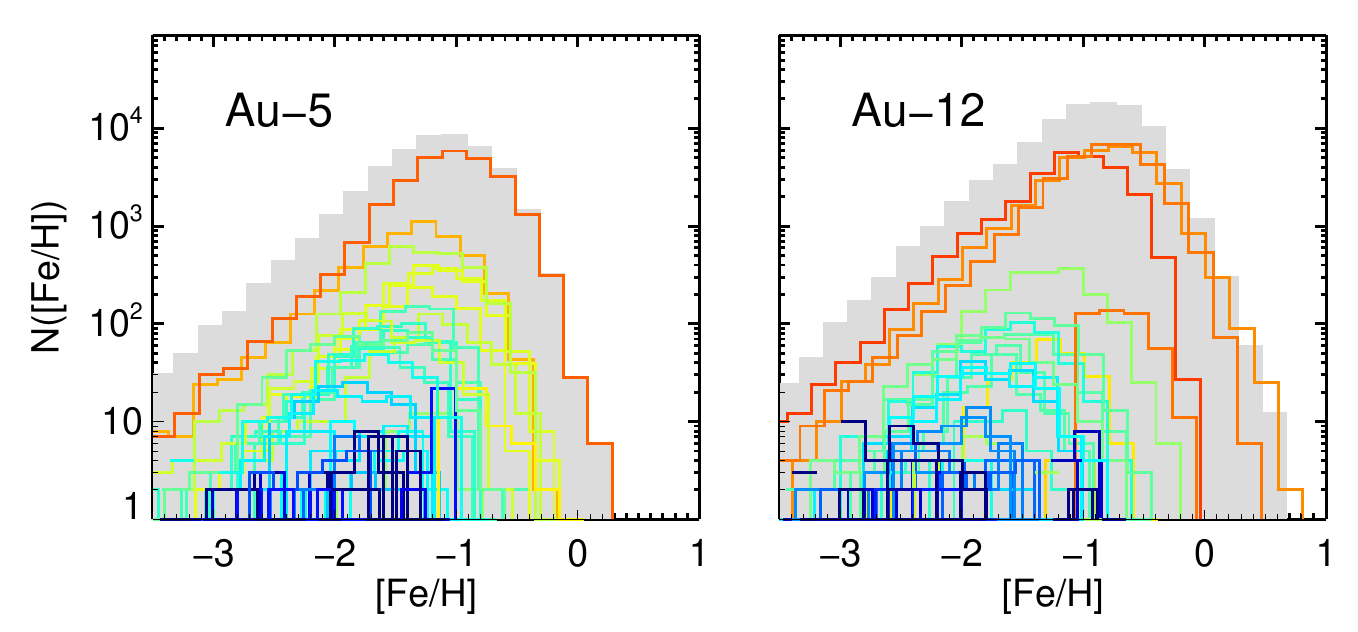}
	\caption{The [Fe/H] distribution of accreted stars within
          galactocentric distance $(20-50)\,$kpc, for two Auriga
          haloes (Au-5 and Au-12 in the left and right panels,
          respectively), subdivided according to their progenitor
          dwarf galaxy.  The distributions are colour-coded by the
          progenitor's stellar mass at infall (colour coding as in
          Fig.~\ref{fig:radialdist}). The total distribution in the
          $(20-50)\,$kpc radial range is shown as a solid histogram in
          the background. The massive dwarfs dominate the mass
          fraction of metals contributed to the stellar halo, even at
          the lowest metallicities.}
    \label{fig:feh_example}
\end{figure}

\begin{figure*}
	\includegraphics[width=17.cm]{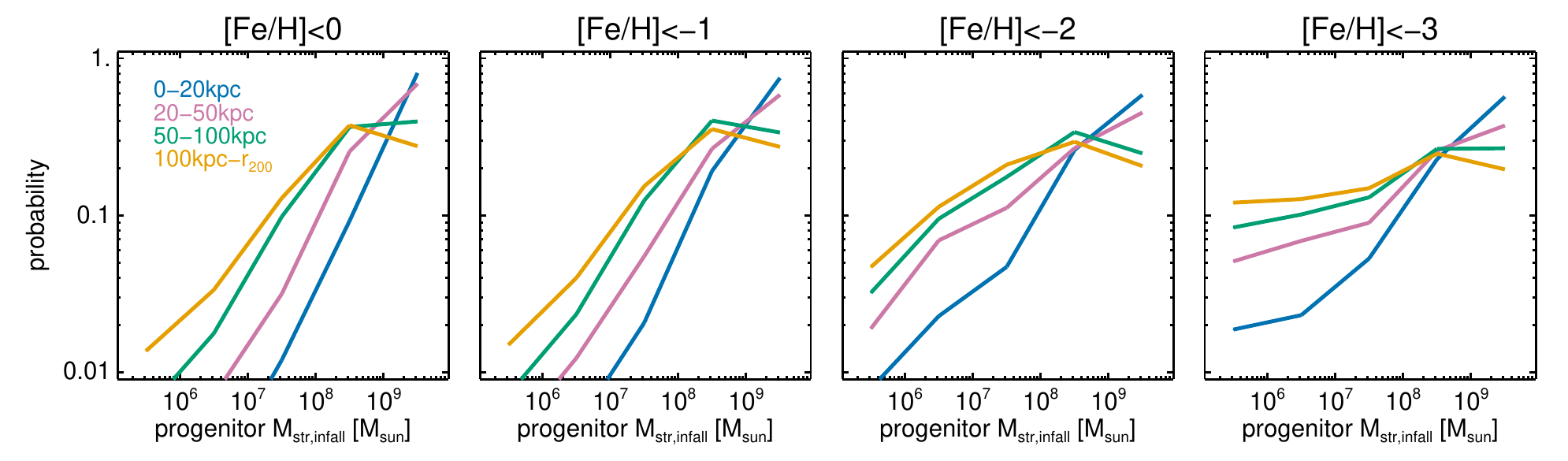}
	\caption{The probability (given by the mass fraction) that
          accreted stars of a given metallicity originated in 
          progenitor dwarf galaxies of different stellar mass. The various panels
          correspond to different [Fe/H] cuts, as stated in the
          legend, and lines of different colour correspond to
          different radial ranges. Even at the lowest metallicities
          ([Fe/H] $<-3$), the most likely contributors to the accreted
          stars are the most massive dwarfs.}
    \label{fig:feh}
\end{figure*}

\subsection{Metallicity of the stellar halo}
\label{sec:halometals}

Lastly, we examine the metallicity contribution of various destroyed
dwarf galaxies to the stellar halo. For the results presented in this
work, the [Fe/H] values of stars have been shifted by roughly
$0.5\,$dex so that the median metallicity in the disc matches that
measured from SDSS in the solar neighbourhood \citep[see][for more
details]{Fattahi2019}. Fig.~\ref{fig:feh_example} illustrates, for two
example haloes (the same ones as in Fig.~\ref{fig:radialdist}), the
[Fe/H] distribution of accreted stars in a spherical shell of
$20-50\,$kpc of the halo. The filled histogram in each panel shows the
total distribution, while individual curves correspond to various
accreted (and destroyed) dwarf galaxies, colour-coded by their
stellar mass at infall. These examples demonstrate that the overall
distribution is dominated by the most massive dwarfs, particularly at
higher metallicities. As the mass of dwarf galaxies decreases, the
peak of their [Fe/H] distribution moves towards lower values, a
reflection of the stellar mass-metallicity relation. However, the
lower mass dwarfs never dominate the overall distribution, even at
metallicities as low as [Fe/H]$\sim -3$ \citep[see also,][]{Deason2016}.
 
We illustrate this quantitatively in Fig.~\ref{fig:feh}, where we
extend the results to all Auriga haloes and various radial
ranges. Each panel shows the probability (mass fraction) that accreted
stars with [Fe/H] lower than a certain value, given in the legend,
originate from progenitors of various stellar masses. Each panel
presents results for four radial ranges. As expected, the higher
metallicity bin ([Fe/H]<0) is dominated by massive dwarf galaxies of
$M_{\rm star}^{\rm infall} > 10^8 \Msun$, at all radii. Interestingly,
dwarf galaxies below a stellar mass
$M_{\rm star}^{\rm infall} < 10^7 \Msun$ only contribute a small
fraction, even to the lowest metallicity bin ([Fe/H]$<-3$); in
particular in the innermost regions which are the most accessible to
observations, less than 10 percent of stars originate from the lower
mass dwarfs. These results have important implications for studies of
metal poor stars in the halo, which are often assumed to come from the
lowest mass dwarfs \citep{Frebel2015}.

\section{Limitations}

\red{In the simulations the contribution of dwarf galaxies to the
  stellar halo depends on the stellar mass-halo mass (SMHM) relation,
  and this in turn depends on the subgrid model, which can differ in
  different simulations. The SMHM relation in the Auriga simulations
  tends to be systematically higher than the relation inferred from
  abundance matching, in the mass range of bright dwarf galaxies,
  $\Mstr \sim 10^8-10^9 \Msun$ \citep{Simpson2018}. We note, however,
  that abundance matching relations have a large scatter in the dwarf
  galaxies regime \citep[see, e.g.,][]{Moster2018,Behroozi2019}. The
  over estimation of stellar mass affects the total stellar halo
  masses but not the systematic trends we found or the relative
  differences between surviving and destroyed satellites. We therefore
  expect that our overall conclusions are not strongly dependent on
  the particular subgrid model of the Auriga simulations but it would
  be desirable to check our results with other simulations of 
  comparable resolution and sample size.}

Other limitations that demand caution concern the lowest metallicity
stars. The Auriga simulations do not resolve ultra-faint dwarf
galaxies and their contribution to the metal-poor end of the stellar
halo metallicity is thus unknown. We anticipate that their
contribution will be small, since their stellar mass is orders of
magnitude smaller than that of the main progenitor of the halo, in
particular in the inner halo. A final point to consider is that the
Auriga galaxy formation models do not include the formation of first
(Population-3) stars, which might affect the shape of the metallicity
distribution at the most metal-poor end.

\section{Summary and conclusions}
\label{sec:summary}

We have used the Auriga $\Lambda$CDM magneto-hydrodynamics simulations
of 28 MW-mass haloes to study the dwarf galaxies that were accreted
onto Galactic haloes after $z\sim3$ ($\tinf>2$ Gyr). We have
considered the population of dwarfs that were destroyed and formed the
accreted stellar halo and contrasted their properties with those of
the dwarfs that survived to $z=0$ and constitute the satellite
population.  The summary of our conclusions is as follows:

\begin{itemize}
    
\item The luminosity function of the total accreted population
  (destroyed + surviving) is similar in all Auriga haloes, with
  little halo-to-halo variation ($0.1\,$dex scatter around the
  mean). This reflects the predictable average assembly history of
  haloes of a given mass in $\Lambda$CDM
  \citep[e.g.][]{Fakhouri2010}.

\item The number of destroyed dwarf galaxies is greater than the
  number of surviving satellites at all stellar masses. Averaged over
  all haloes, $30$ percent of accreted dwarfs with infall stellar
  mass, $M_{\rm star}>10^5 \Msun$, survive to $z=0$. This does not
  necessarily imply that the ex-situ stellar mass of the halo is
  larger than the combined mass of the surviving satellites since a
  significant fraction of destroyed dwarf debris is deposited in 
  the disc and the very central regions of the haloes.


\item The fraction of surviving dwarf galaxies is strongly dependent
  on infall time as well as on the mass at infall. Roughly 90~percent
  of the dwarfs that were accreted later than $\tinf = 10\,$Gyr \red{(present day corresponds to $13.8\,$Gyr)} survive to
  the present; this fraction drops to only 15~percent for dwarfs that
  were accreted early ($\tinf = 2-6\,$Gyr). The survival fraction also
  depends on the mass at infall, such that both high mass and low mass
  dwarfs are destroyed more efficiently while accreted dwarfs of
  $M_{\rm star, infall} \sim 10^7 \Msun$ are the most
  resilient. 

\item The average infall time of surviving satellites is
  $\sim 7\,$Gyr, with a dependence on the mass at infall. For
  satellites of $M_{\rm star,infall} \sim 10^6 \Msun$ and
  $10^9 \Msun$, the typical infall times are $\sim6\,$Gyr and
  $\sim10\,$Gyr respectively. In contrast, destroyed dwarfs have an
  average infall time, $\tinf \sim 2.5$ and $\sim5\,$Gyr, for
  $M_{\rm star,infall}\sim 10^6$ and $10^9\Msun$, 
  respectively. 

\item Due to the slight evolution of the stellar mass-metallicity
  relation with time, the later infall of surviving satellites results
  in their metallicity ([Fe/H]) at infall being higher than that of
  their destroyed counterparts of similar stellar mass. This
  difference is 0.2~dex at stellar mass
  $M_{\rm star,infall}\sim 10^8-10^9 \Msun$. The higher mass
  satellites continue forming stars after infall, causing their [Fe/H]
  to increase further by $\sim 0.2\,$dex to the present. 


\end{itemize}

In the second part of the paper, we focused on the material deposited
in the main galaxy by destroyed (and disrupted) dwarfs and on the
assembly of their accreted stellar haloes. These results are
complementary to those of \citet{Monachesi2019}. We have extended the
results of that work by examining progenitors of the accreted stellar
halo as a function of radius. We also studied additional properties such as
the dark matter and metallicity contributions of the destroyed
dwarfs. 

We identified all stars that formed later than $z=3$ ($t=2\,$Gyr) in
progenitors that are not the main progenitor of the central galaxy, but
are bound to it at
$z=0$. 
Our results concerning this accreted component are: 

\begin{itemize}

\item In agreement with \citet{Monachesi2019}, we find that the total
  accreted mass of the stellar halo within $r_{200}$ is brought in by
  a few relatively massive dwarf galaxies. However, we show that the
  number varies as a function of galactocentric radius ($r_{\rm
    GC})$. The innermost regions, $r_{\rm GC}<20\,$kpc, have typically
  $\sim 3$ significant dwarf progenitors which make up 90~percent of
  the mass; this number changes to $5$ and $8$ for
  $r_{\rm GC}=(20-50)\,$kpc and $r_{\rm GC}>50\,$kpc respectively.

\item In the inner 20\,kpc, the contribution of individual dwarfs
  drops rapidly with the mass of the dwarf, such that more than
  50~percent of the mass typically comes from a single massive galaxy
  ($M_{\rm star}>10^8 \Msun$). In the outer regions, the contribution
  from various progenitors is comparable. The recent discovery of the
  {\it Gaia-sausage-Enceladus} component
  \citep{Helmi2018,Belokurov2018} is consistent with our
  findings. Indeed, this merger is thought to dominate the mass of the
  inner Galactic stellar halo.

\item The contribution of debris from existing satellites is
  significant in the outer regions of the stellar haloes. On average,
  $\sim30\pm25$ percent of the ex-situ mass in the
  $r_{\rm GC}=100\,$kpc$-r_{200}$ shell is made up of such stars. This
  fraction is much smaller, $<1\%$, in the inner $20\,$kpc region.


\item The significant contributions of more massive dwarf galaxies to
  the galactic stellar haloes have important implications for the
  metallicity content of the halo. Unsurprisingly, the more massive
  destroyed dwarfs are the main origin of high metallicity
  ([Fe/H]$\gtrsim -1$) stars. However, we find that even at the more
  metal-poor tail of the distribution stars originating in low mass
  dwarf galaxies never dominate. In the inner $20\,$kpc, stars more
  metal poor than [Fe/H]$<-3$ have less than 10~percent chance, on
  average, of having been deposited by dwarf galaxies of mass less than
  $\Mstr=10^7\Msun$.
  
\item We show that the accreted stellar haloes are formed inside out;
  i.e. the top progenitors of the inner stellar haloes have \red{median infall
  time of $\tinf\sim4\,$Gyr after the Big Bang, whereas the top
  progenitors of the outer 100 kpc fell in on average at $\tinf\sim7\,$Gyr}. 

\item We show that the build-up of the stellar halo is significantly
  different from that of the dark matter. The dark matter contribution
  from the stellar halo progenitors is typically an order of magnitude
  lower than their stellar contribution inside $r_{200}$.  
  This result is a consequence of the steep
  stellar mass-halo mass relation at the low mass end, and the fact
  that a significant fraction of dark matter halo mass is built up
  from dark subhaloes and smooth accretion. The radial differences are
  due to differences in the tidal stripping of dark matter and  stars
  from accreted dwarf
  galaxies. 

\end{itemize}

Our findings regarding both surviving satellites and destroyed dwarf
galaxies indicate that the observed satellites at the present day are
{\it not} the building blocks of the stellar halo. The building blocks
of the stellar halo are a biased population of dwarf galaxies which
fell in relatively early, and differ from the observed satellites in
particular in their metallicity content. The metallicity of the
stellar halo (which is formed from the destroyed dwarfs) is predicted
to be lower than that of the existing satellites, which is consistent
with observations of MW and Andromeda
\citep{Vargas2014,Escala2020,Kirby2020}.

While the observed number of satellites is dominated by low mass
dwarfs, low mass dwarf galaxies contribute negligibly to the mass of
the inner stellar halo, even at low metallicities. This implies that
the kinematic properties of the metal poor stars are biased and differ
from what is expected from the accretion of numerous dwarf galaxies
with a variety of orbital parameters. This will be the subject of
future work.

\section*{Acknowledgements}

We are thankful to the anonymous referee for their helpful feedback. AF is supported by an Marie-Curie COFUND/Durham Junior Research
Fellowship (under EU grant agreement no. 609412), and AD by a Royal
Society University Research Fellowship. AD, AF and CSF are also
supported by the Science and Technology Facilities Council (STFC)
[grant numbers ST/F001166/1, ST/I00162X/1,ST/P000541/1]. CSF is also
supported by ERC Advanced Investigator grant, DMIDAS [GA 786910]. FM
acknowledges support through the Program `Rita Levi Montalcini' of the
Italian MIUR. AM acknowledges financial support from CONICYT FONDECYT
Regular 1181797. FAG acknowledges financial support from CONICYT
through the project FONDECYT Regular Nr. 1181264. FAG and AM
acknowledge funding from the Max Planck Society through a Partner
Group grant.

This work used the DiRAC Data Centric system at Durham University,
operated by the ICC on behalf of the STFC DiRAC HPC Facility
(www.dirac.ac.uk). This equipment was funded by BIS National
E-infrastructure capital grant ST/K00042X/1, STFC capital grant
ST/H008519/1, and STFC DiRAC Operations grant ST/K003267/1 and Durham
University. DiRAC is part of the National E-Infrastructure. 

\section*{Data Availability}

The data presented in figures will be shared on reasonable request to the corresponding author. Raw simulation data can be shared on reasonable request to the Auriga team \citep{Grand2017}.




\bibliographystyle{mnras}
\bibliography{master} 




\appendix

\section{Convergence}
\label{sec:convergence}

\begin{figure*}
	\includegraphics[width=17.cm]{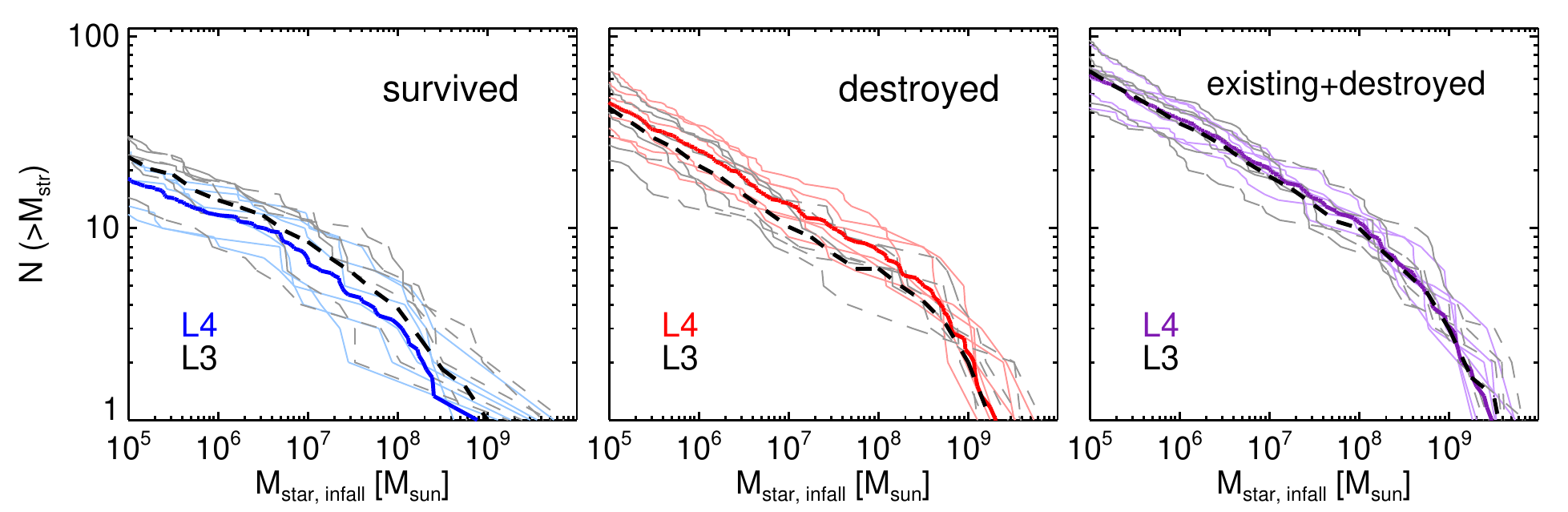}
	\caption{Similar to Fig.~\ref{fig:SMF} but showing 
          convergence between the higher resolution (L3) and the
          fiducial resolution (L4) simulations. 
          Only the 6 haloes that were simulated at both L3 and L4
          resolution are considered. The grey dashed and black dashed curves correspond to the L3 results, while the colour lines show the L4 results. }
	\label{fig:SMF_conv}
\end{figure*}

\begin{figure*}
	\includegraphics[width=15.cm]{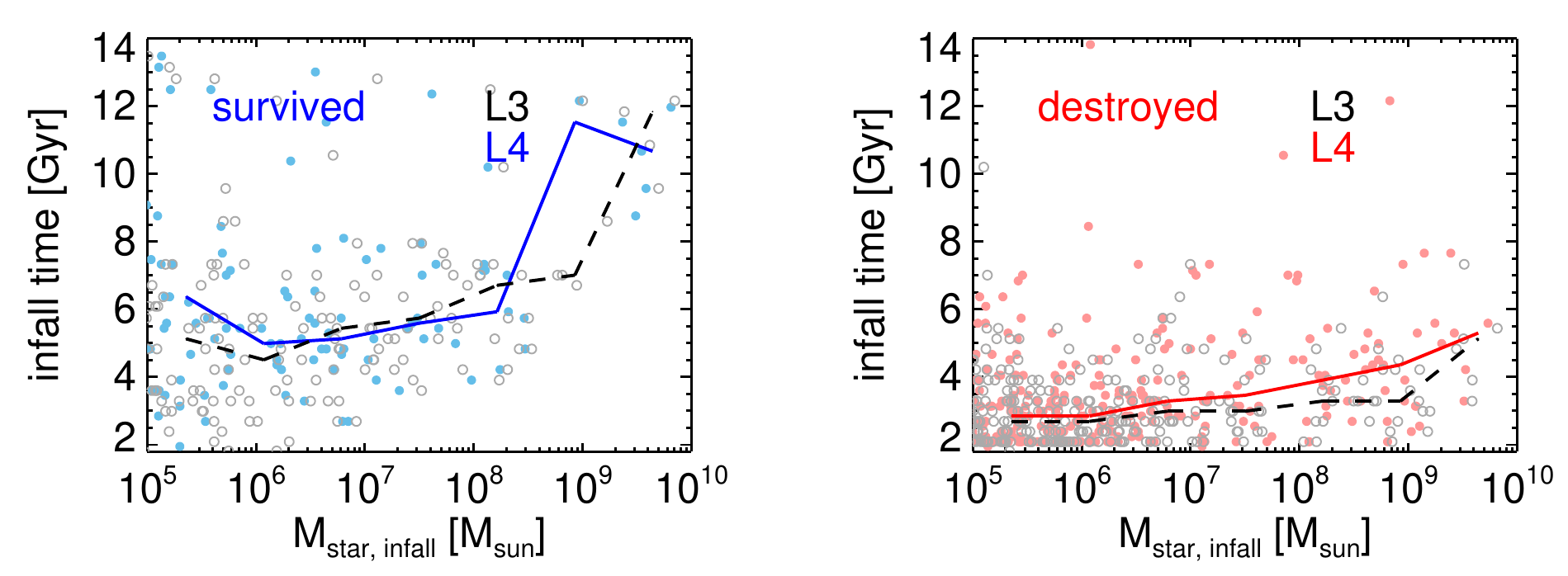}
	\caption{Similar to Fig.~\ref{fig:tinfall} but divided into
          surviving and destroyed dwarfs in the left and right panels,
          respectively. Small grey and colour symbols illustrate
          individual dwarfs at L3 and L4 resolution
          respectively. Similarly, black and colour curves show the
          average for the two resolutions.}
    \label{fig:tinfall_conv}
\end{figure*}

\begin{figure}
    \hspace{-0.2cm}
	\includegraphics[width=8.2cm]{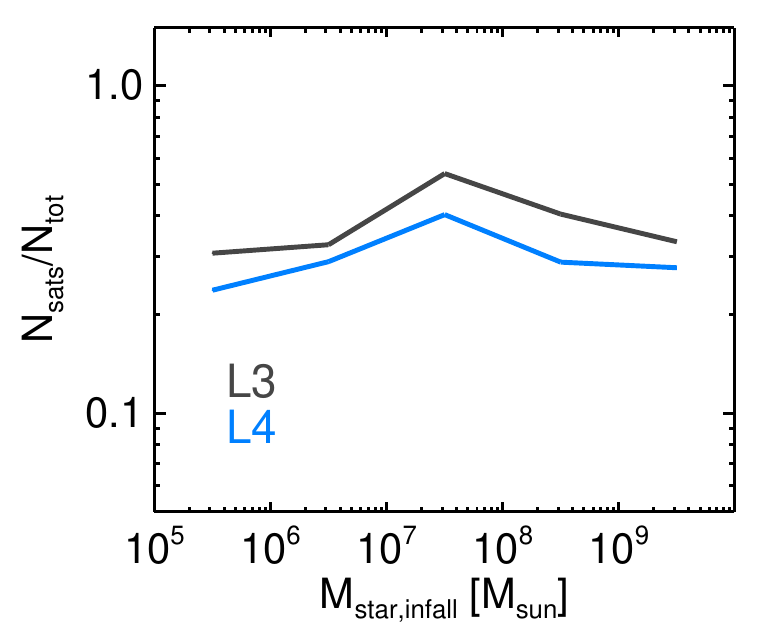}
	\caption{The fraction of surviving dwarfs (satellites),
          relative to all accreted dwarfs, as a function of stellar
          mass at infall for the six Auriga haloes at resolution
          levels L3 and L4. Because of its small size we do not split
          the sample into bins of infall time.}
    \label{fig:fraction_conv}
\end{figure}

\begin{figure}
    \hspace{-0.2cm}
	\includegraphics[width=8.2cm]{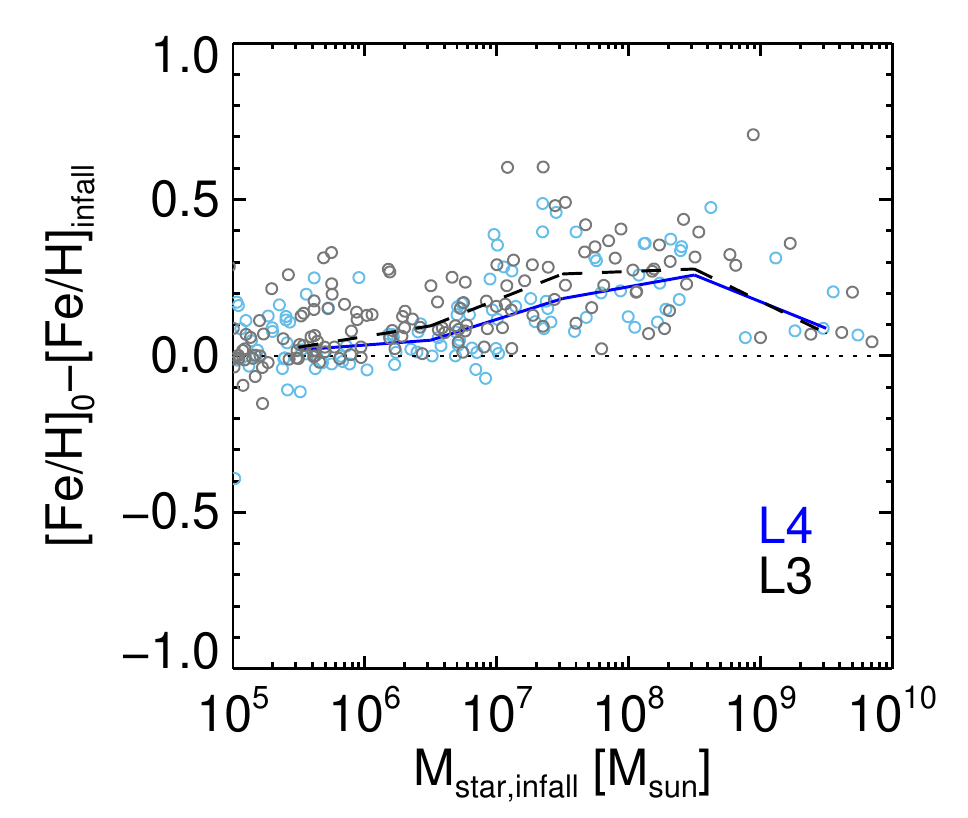}
	\caption{\red{Similar to the bottom-right panel of Fig.~\ref{fig:z_mstar} but showing the convergence between L3 and L4 runs. Small grey and coloured symbols illustrate individual dwarfs at L3 and L4 runs, respectively. Similarly, black and colour curves show the average for the two resolutions.}}
    \label{fig:deltaz_conv}
\end{figure}

We quantify the degree of convergence between the L4 and L3 Auriga
simulations in Figs.~\ref{fig:SMF_conv}, \ref{fig:tinfall_conv}, and
\ref{fig:fraction_conv} which are the equivalent of
Figs.~\ref{fig:SMF}, \ref{fig:tinfall}, and \ref{fig:fraction} in the
main text.  \red{Fig.~\ref{fig:deltaz_conv} is equivalent to the
  bottom-right panel of Fig.~\ref{fig:z_mstar}.}  All these
convergence plots include only the 6 haloes that were simulated at
both resolutions.

The luminosity function of the total accreted population (right panel
of Fig.~\ref{fig:SMF_conv}) shows excellent convergence between the
two resolution simulations. However, we note that $\sim 10$ percent of
dwarfs have moved from the destroyed population to the surviving one.

The definition of ``destroyed'' in the L3 simulations presented here
is the same as in the L4 resolution simulation: a halo mass less than
$10^7\Msun$ or a stellar mass, $\Mstr < 10^5 \Msun$. We tried relaxing
this definition to include any accreted dwarf that was completely
destroyed by $z=0$, and the result did not change in any meaningful
way. We note that dwarf galaxies form in relatively massive,
well-resolved haloes ($10^9-10^{10})\Msun$ haloes or $(10^4-10^5)$
particles in L4), not in haloes at the resolution limit.

As implied in Fig.~\ref{fig:SMF_conv}, dwarf galaxies in the higher
resolution runs have a $\sim 10$ percent higher chance of
surviving. This is independent of stellar mass, and indicates that the
downturn in the surviving fraction at lower masses is not a resolution
effect.

 

\bsp	
\label{lastpage}
\end{document}